\documentclass[aps, twocolumn, superscriptaddress]{revtex4-1}

\usepackage{graphicx}
\usepackage{amsmath}
\usepackage{dsfont}
\usepackage{amssymb}
\usepackage{physics}
\usepackage{hyperref}
\usepackage{ulem}
\hypersetup{colorlinks=true,
	    final=true,
	    linkcolor=blue,
	    citecolor=blue,
	    filecolor=blue,
	    urlcolor=blue,}

\newcommand{\beq}{\begin{equation}}
\newcommand{\eeq}{\end{equation}}

\usepackage{amsmath,amsfonts,amssymb}
\usepackage{graphicx}
\usepackage{setspace}
\usepackage{tocloft}

\begin{document} 

\title{Homemade kit for demonstrating Barkhausen Effect}

\author{Shantanu Shakya}
\affiliation{Department of Physics, Indian Institute of Space Science and Technology(IIST), Thiruvananthapuram, Kerala - 695547}

\author{Navinder Singh}
\affiliation{Theoretical Physics Division, Physical Research Laboratory,  Navrangpura, Ahmedabad - 380009}

\renewcommand{\cftdotsep}{\cftnodots}
\cftpagenumbersoff{figure}
\cftpagenumbersoff{table}

\begin{abstract}
This paper presents an innovative and cost-effective approach to understanding the Barkhausen effect through the design and implementation of an educational kit. The Barkhausen effect, characterized by Barkhausen noise (BN) during magnetization changes in soft magnetic materials, is explored for its application in probing hysteresis properties and magnetization dynamics. The study investigates scaling properties, categorizing ferromagnetic materials based on scaling exponents. The primary contribution is the introduction of a practical and accessible kit for hands-on Barkhausen Effect demonstrations, revolutionizing the educational experience. This kit enables students to not only comprehend the intricacies of BN but also calculate the scaling constant ($\tau$) for Soft Iron samples. The paper demonstrates the successful construction of the kit, its signal amplification capabilities, and data collection accuracy, showcasing its potential for widespread educational use.
\end{abstract}

% Include a list of up to six keywords after the abstract
\keywords{Barkhausen Effect, Magnetization Dynamics, Domain Wall Motion, Scaling Properties, Ferromagnetic Materials, Educational Kit, Soft Iron, Scaling Constant}

% Include email contact information for corresponding author
%{\noindent \footnotesize\textbf{*}Fourth author name,  %\linkable{myemail@university.edu} }

\begin{spacing}{1.1}   % use double spacing for rest of manuscript

\maketitle

\section{Introduction}
\label{sect:intro}  % \label{} allows reference to this section
\begin{figure}[h!]
    \centering
    \includegraphics[scale = 0.25]{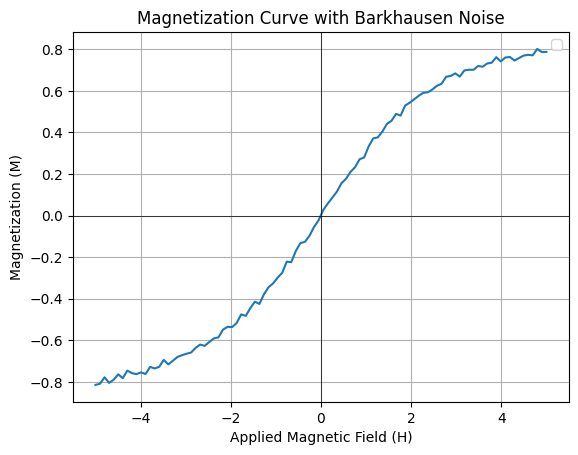}
    \caption{Barkhuasen Noise in magnetization curve}
    \label{fig:BN_noise}
\end{figure}

Barkhausen in 1919 noticed that changes in the magnetization of an iron sample induced a special form of time dependent electric potential in a pickup coil wound over the sample. Upon amplification, this led to the generation of audible "noise" as depicted in Figure \ref{fig:BN_noise}. This phenomenon is widely known as "Barkhausen noise" \cite{Feynman}. Barkhausen suggested its association with irreversible alterations in magnetization.
When exposed to a magnetic field undergoing a gradual temporal shift, such as the slow movement of a manually-driven magnetic yoke, a usual magnetic material like iron produces irregular signal in a pick-up coil. This irregular acoustic manifestation contrasts sharply with the systematic pattern of the applied magnetic field and is detectable through a microphone. This marked the initial indirect confirmation of the presence of magnetic domains, a concept theorized a few years earlier by Weiss \cite{Weiss}.\\

Soon after its discovery, it became apparent that Barkhausen noise (BN) could serve as a valuable tool for delving into and understanding the magnetization dynamics inherent in soft magnetic materials, providing insights into specific hysteresis properties. While the measurement of BN is relatively straightforward, its interpretation presents a greater challenge. This complexity stems from the stochastic nature of domain wall (DW) motion, which progresses through sudden jumps or avalanches during slow magnetization. These dynamics are notably influenced by material microstructure, along with additional factors such as the demagnetizing field, external stress, and others \cite{Durin}.

\section{Experimental Idea}

In this section, we will delve into the details of the experiment, including the instruments employed, and the associated circuitry. We will explore both the auditory aspect of detecting Barkhausen noise (BN) and the methods for its measurement.\\

\begin{figure}[h!]
    \centering
    \includegraphics[scale = 0.2]{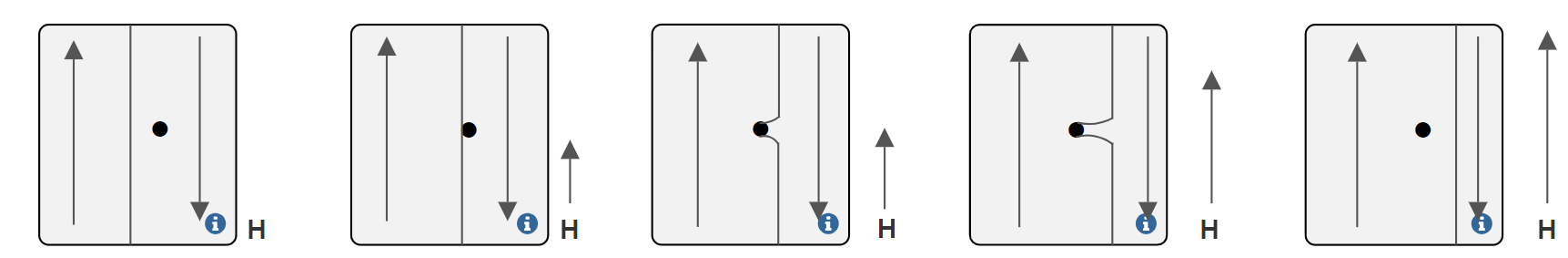}
    \caption{Pinning of domain wall due to impurities}
    \label{fig:Wall motion}
\end{figure}

BN is generated by alterations in the magnetization of a ferromagnetic material. This change in magnetization is often attributed to the presence of impurities and defects within the material, which can act as obstacles, pinning the movement of domain walls (as shown in figure \ref{fig:Wall motion}). When we apply an external magnetic field and increase it beyond a critical value denoted as $H_c$, a pivotal event known as the "depinning transition" occurs. During this transition, the pinned domain walls are suddenly released or depinned, resulting in a significant jump in the material's magnetization curve. This jump can be detected using a pickup coil that operates on the principles of "Faraday's law of electromagnetic induction".\\

According to Faraday's law, any alteration in magnetic flux within the coil's vicinity induces a corresponding voltage at the coil's terminals. This induced voltage is the key to our ability to both hear and measure the Barkhausen noise in our experiment.\\
\begin{equation}
    V=N*A*\frac{dB}{dT}
\end{equation}
V- voltage induced in the coil\\
N- number of the turns of the pick-up coil\\
A- cross section area of the coil\\
B- magnetic field\\

When deliberately and gradually adjusting the magnetic field strength, either increasing or decreasing, the coil induces a prominent voltage signal characterized by a large amplitude and low frequency. In contrast, the rapid Barkhausen events, characterized by much smaller amplitudes, contribute to a signal with higher frequency. This high-frequency signal is superimposed upon the slower and larger signal. Our aim is to filter out the BN noise voltage and reject the slower and larger signals.\\

\begin{figure}[h!]
    \centering
    \includegraphics[scale = 0.3]{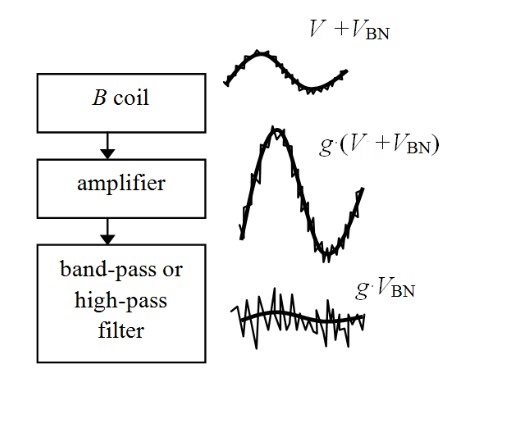}
    \caption{This is how we are extracting the signal\cite{BooktMeaseurement}}
    \label{experiment idea}
\end{figure}

Figure 1 of reference 4\cite{thinFilmm} illustrates the magnetization curve of a NiFe film \cite{thinFilmm}. Initially appearing smooth when recorded under the influence of a slowly varying external magnetic field, a closer inspection with magnification reveals non-uniform changes in magnetization. This non-uniformity is marked by the presence of high-frequency, low-amplitude Barkhausen noise, which is superimposed on the main signal.\\
%\begin{figure}[H]
    %\centering
    %\includegraphics[scale = 0.5]{NiFe pickup coil.jpg}
    %\caption{Magnetization curve, as a function of the time,of a 100-nm-thick ferromagnetic %NiFe film submitted to a smooth, slow-varying external magnetic field.\cite{thinFilmm}}
   % \label{fig:NiFe}
%\end{figure}
To extract the Barkhausen noise from the recorded data, we adopt a two-step approach. Initially, a high-gain amplifier is employed to boost the amplitude of the signal. Following this amplification, we implement suitable filtering techniques to isolate the Barkhausen noise. This is accomplished using band-pass or high-pass filters, effectively eliminating lower-frequency components associated with slower and larger signals (refer to Figure \ref{experiment idea}). Frequencies below 300 Hz are filtered out.\cite{thinFilmm}.

This approach enables us to discern and analyze the Barkhausen noise, a phenomenon of particular interest for further investigation and understanding.\\

\section{Experimental set-up and Proposal of Kit}
\label{sect: experi}
To conduct this experiment, we devised a straightforward homemade kit, from which we extracted the Barkhausen Noise. The kit comprises the following components:
\begin{itemize}
    \item Ferromagnetic Sample
    \item Pick-up coil
    \item Speaker
    \item Permanent Magnet
    \item Audio Amplifier
\end{itemize}
\subsection{Ferromagnetic Sample}
In this experiment, we are utilizing a cylindrical soft iron core with a diameter of 19.5 mm and a length of 71 mm. Soft iron is an easily accessible ferromagnetic material characterized by high permeability and low coercivity, meaning it does not retain its magnetization when the magnetic field is removed. This property allows for easy magnetization and demagnetization, and it exhibits minimal hysteresis loss, making it the ideal material for our experiment, where we require repeated switching of the magnetic field to accurately observe Barkhausen noise (BN). \\

In soft Iron core of not very high purity which is used in standard applications such as in concrete structures etc. These impurities are the primary cause of Barkhausen Noise. Non-magnetic impurities, such as carbon, sulfur, phosphorous, and silicon, are present, along with magnetic impurities like nickel, cobalt, and manganese. Additionally, the material contains grain boundary defects, dislocations within its crystal structure (which can impede the movement of domain walls), as well as inclusions and surface defects.

\subsection{Pick-up Coil}
The pickup coil serves the crucial function of detecting and converting alterations in magnetic flux into electrical signals. It plays a pivotal role in various devices where the conversion of magnetic fields into electrical signals is essential for purposes such as detection, measurement, or transmission. The operation of a pickup coil is based on Faraday's law of electromagnetic induction. The magnitude of the voltage signal it produces depends on several factors, including the strength of the magnetic field, the number of turns within the pickup coil, and the rate at which the magnetic field changes.\\

For our experiment, we have specifically designed a four-layered pickup coil using enameled copper wire with a total of 300 turns. The copper wire employed in this coil has a cross-sectional diameter of 0.5 mm. This coil is wound around the sample (Soft Iron core), enabling it to effectively capture the Barkhausen noise that we are studying in our experiment.

\begin{figure}[h!]
        \centering
        \includegraphics[scale = 0.3]{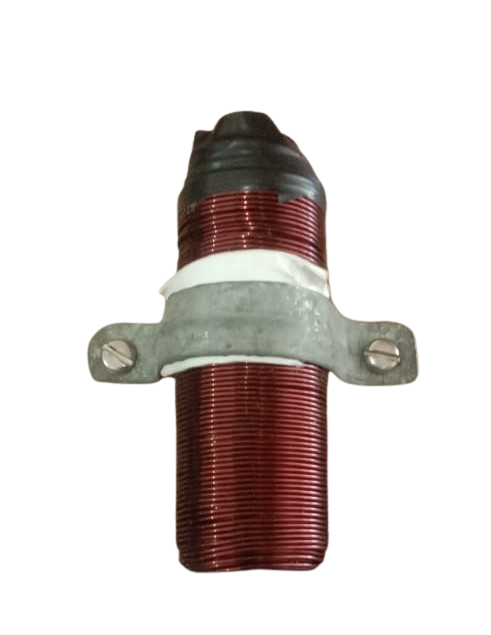}
        \caption{Pickup coil wound over soft iron (sample)}
        \label{fig:enter-label}
    \end{figure}
    
\subsection{Speaker}
A speaker is an electroacoustic transducer designed to transform electrical signals into sound waves. It functions through the interaction between a permanent magnet and an electromagnet that are affixed to a diaphragm. This interaction results in the diaphragm's movement, which corresponds to the incoming electrical signal, ultimately producing sound.\\

In our experimental setup, crafted for demonstration purposes, we have incorporated an 8 $\Omega$ impedance speaker with a power rating of 0.25 W. This speaker enables us to audibly perceive the faint yet distinct Barkhausen noise (BN) when one listens closely to it.

\begin{figure}[h!]
     \centering
     \includegraphics[scale = 0.25]{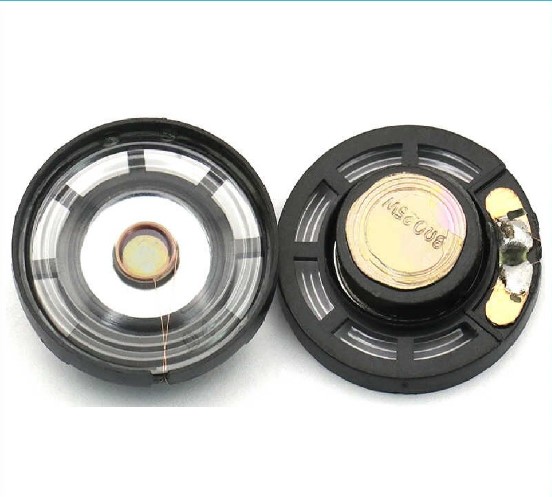}
     \caption{speaker used for this experiment}
     \label{fig:enter-label}
 \end{figure}

\subsection{Magnet}
Certainly, for this demonstration, an external magnetic field is essential for the sample. We are utilizing a "ring-shaped" permanent ferrite magnet, which we retrieved from old speakers. This ferrite magnet can generate a magnetic field with a range spanning from a few milli-tesla (mT) to sub tesla (T), depending on orientation and distance from the magnet.\\

Ferrite magnets are the most commonly used type of magnets in speakers. They are typically composed of a mixture of iron oxide and other materials, often strontium or barium. These magnets are not only cost-effective but also possess favorable magnetic properties.\\

When we bring the magnet in proximity to the sample, the sample becomes magnetized, and this magnetization process is accompanied by the generation of Barkhausen noise, which we can audibly perceive.\\

\begin{figure}[h!]
    \centering
    \includegraphics[scale = 0.25]{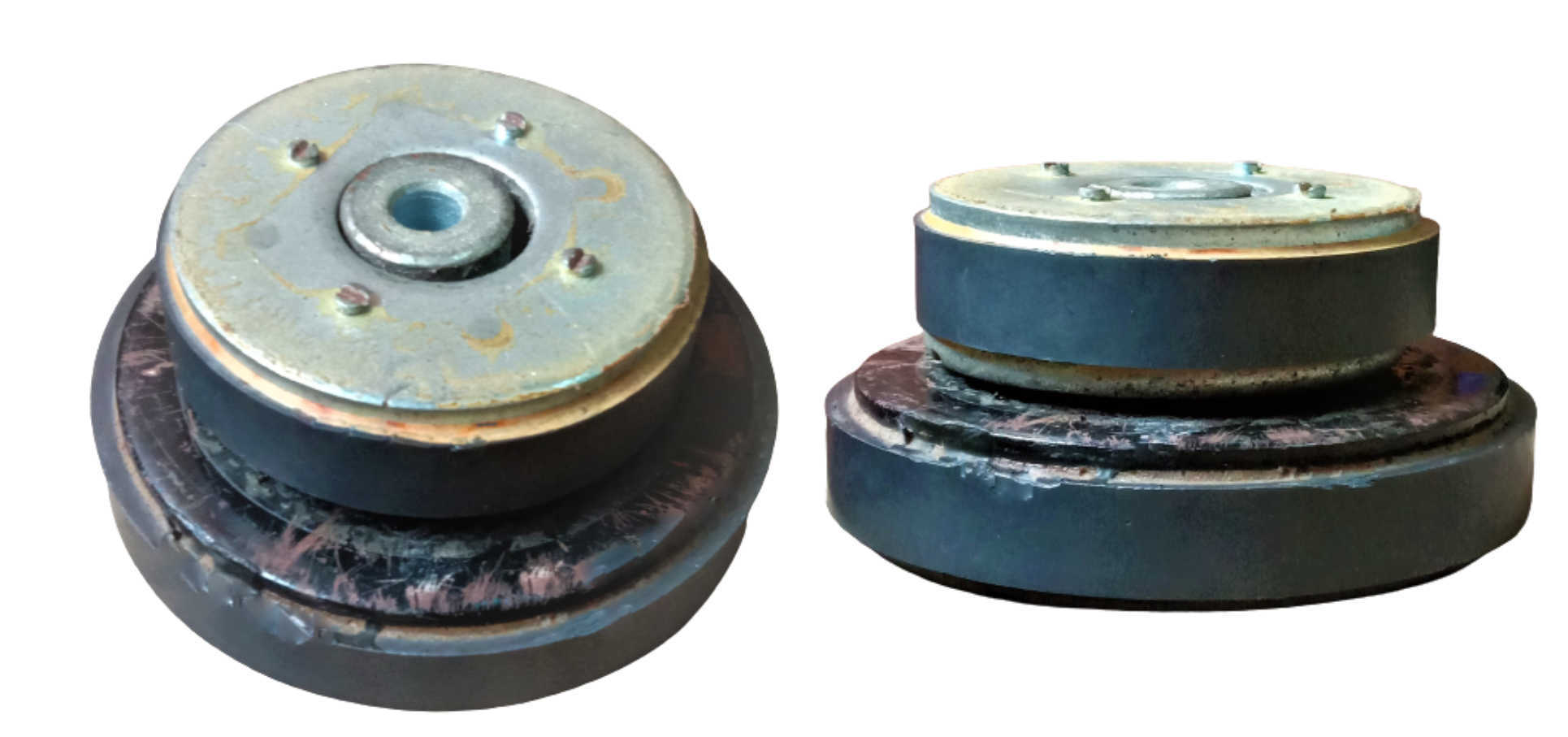}
    \caption{Magnets used.}
    \label{fig:comMag}
\end{figure}

We are using two ferrite ring magnets for our demonstration. The larger one has a diameter of 110 mm and a height of 15mm, while the smaller one has a diameter of 80 mm and a height of 14mm. These magnets are combined and positioned as illustrated in the figure \ref{fig:comMag} for our experiment.

\subsection{Audio Amplifier}
An audio amplifier is an electronic device designed to increase the amplitude of low-power electronic audio signals within the frequency range of approximately 20 to 20,000 Hz. These signals can originate from various sources, such as a radio receiver or an electric guitar pickup, and the amplifier's purpose is to elevate them to a level suitable for driving speakers or headphones.\\

To amplify the Barkhausen noise (BN) in our experiment, we have developed a cost-effective "Audio Amplifier" utilizing the LM386 integrated circuit (IC). This IC offers adjustable gain and boasts low power consumption. We are able to power this amplifier efficiently using a 9V battery.

\subsubsection{LM386}
\begin{figure}[h!]
    \centering
    \includegraphics[scale = 0.3]{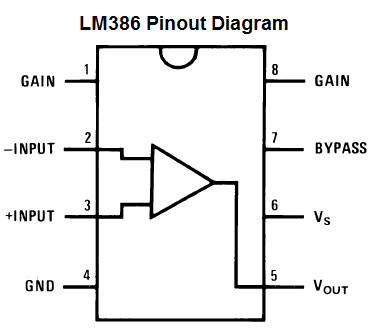}
    \caption{LM386. For details refer to: www.learningaboutelectronics.com/Articles/How-to-connect-a-LM386-audio-amplifier-chip}
    \label{fig:pinout}
\end{figure}

The \textbf{LM386} is a 8 pin IC, used for designing power amplifier which consume very low power, have adjustable gain of any value from 20 to 200, which could be done by connecting appropriate capacitor between pin 1 and 8 (see Figure \ref{fig:pinout}) \cite{LM386}.\\
*For more information, see reference on IC LM386\cite{LM386}
\subsection{Amplifier Circuit}
\begin{figure}
    \centering
    \includegraphics[scale = 0.25]{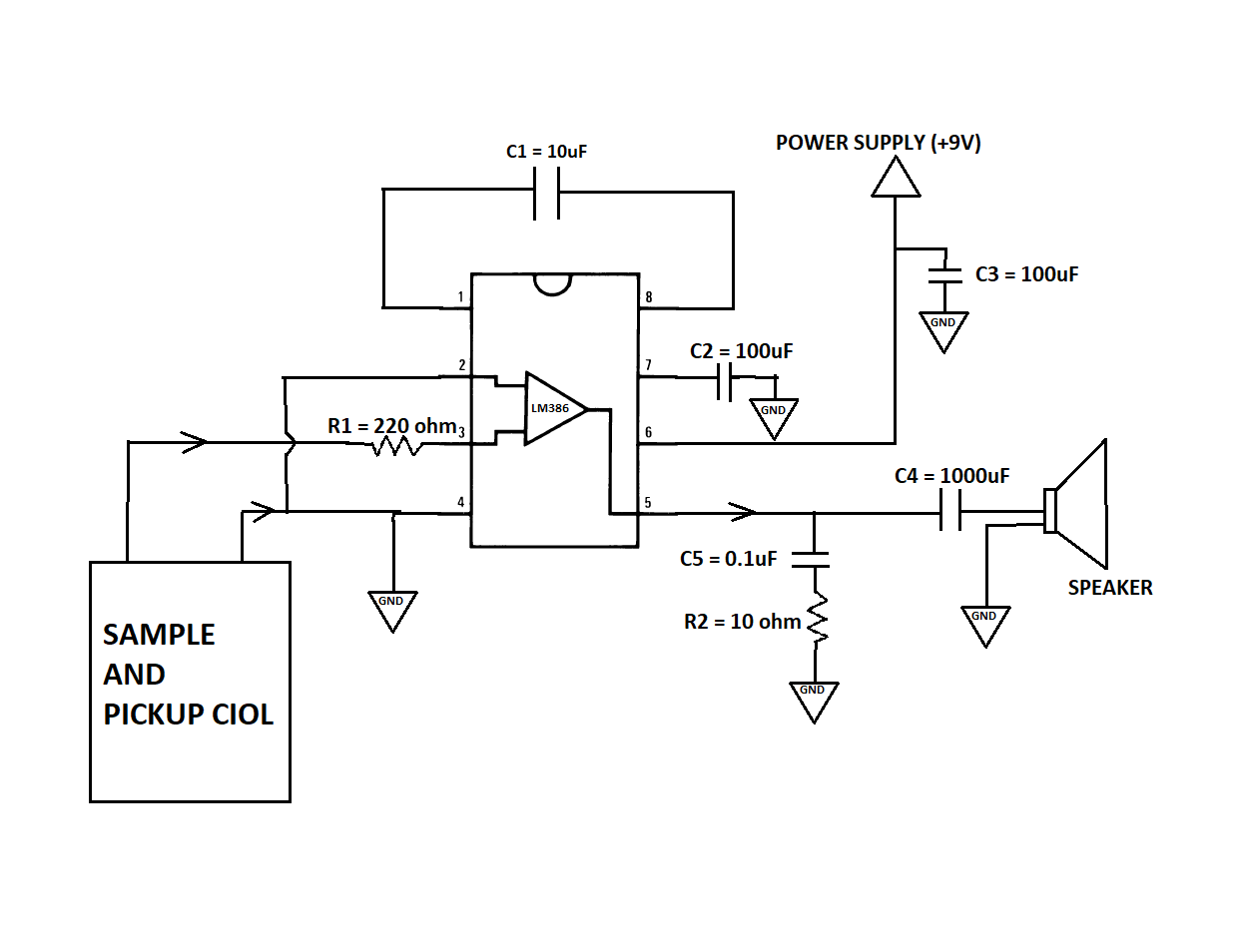}
    \caption{Circuit designed for this kit}
    \label{fig:ampCircuit}
\end{figure}

Figure \ref{fig:ampCircuit} illustrates the amplifier circuit constructed using the LM386 IC. This circuit is designed to enhance the Barkhausen noise received from the pickup coil, with pin 3 serving as the input. To optimize the amplification process, a 220$\Omega$ resistor is connected in series with pin 3. Additionally, a 10$\mu$F capacitor is positioned between pins 1 and 8 to achieve maximum gain. A "bypass" capacitor with a value of 100$\mu$F is connected through pin 7, and another 100$\mu$F capacitor is linked to the power supply to ground any AC components within the power supply.\\

At the output, which is pin 5, a 1000$\mu$F capacitor is utilized to block the DC signal (by the motion of hands) from reaching the speaker. Furthermore, a low-pass filter is incorporated in this section to eliminate any high-frequency noise (specially MHz noise), particularly in the megahertz range, that may be present along with the BN signal.

\begin{figure}
    \centering
    \includegraphics[scale = 0.3]{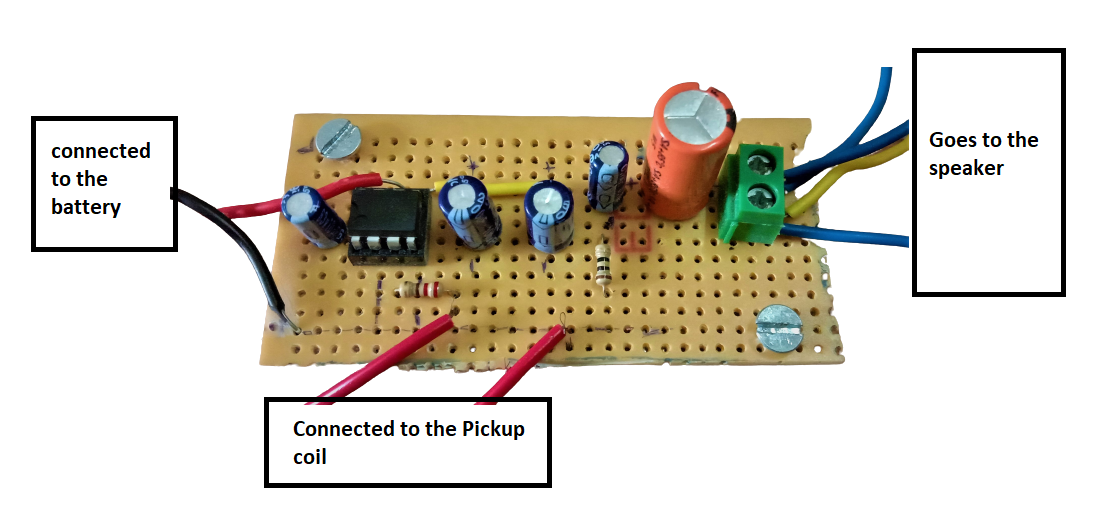}
    \caption{Circuit made on general purpose PCB }
    \label{fig:kitCir}
\end{figure}
\begin{figure}[h!]
    \centering
    \includegraphics[scale = 0.3]{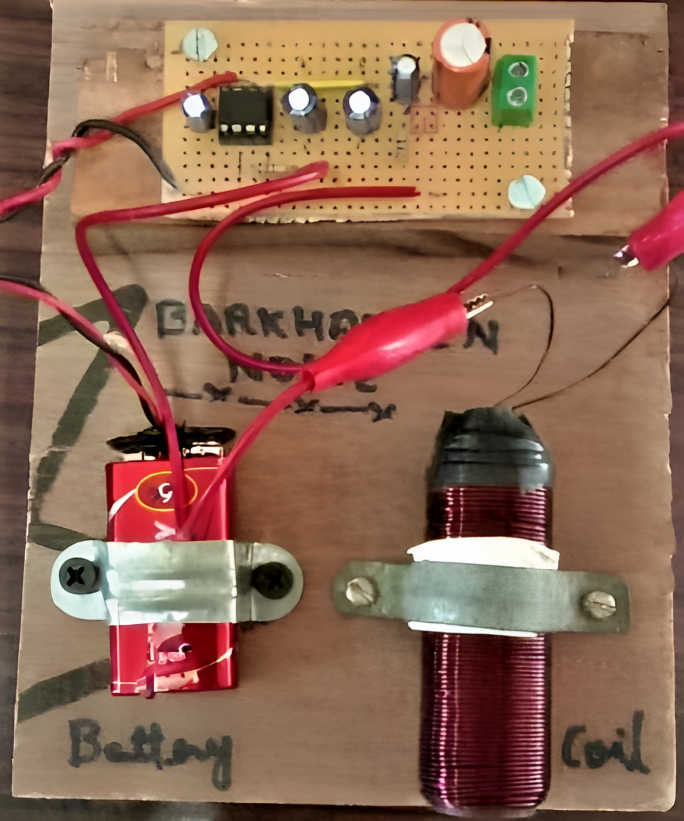}
    \caption{Barkhausen Experiment kit designed by us.}
    \label{fig:complete kit}
\end{figure}

Components are mounted on a piece of plywood, and made a very simple kit. ( see figure \ref{fig:complete kit})\\

It turns out that the total cost to make it is only around Rs 300/-.

\section{Results}
Initially, we successfully constructed a cost-effective kit (refer to Figure \ref{fig:complete kit}), utilizing components detailed in Section \ref{sect: experi}. This kit, as demonstrated in a relevant video accessible as supplementary material at \href{https://tinyurl.com/iistBNdata}{https://tinyurl.com/BNdata}, effectively captures the Barkhausen Noise, producing clear auditory signals through the integrated speaker. Subsequently, we directed the resulting signal to the \textit{Keithley 2450 sourcemeter}, a versatile instrument serving as a digital multimeter equipped with built-in ammeter, voltmeter, and ohmmeter functionalities. Notably, this instrument boasts a remarkable resolution of 10 nV for the voltmeter and 10 fA for the ammeter, featuring 6.5 significant digits and a reading capability ranging from 1 reading per second to 1700 readings per second.\\

\begin{figure}[h!]
    \centering
    \includegraphics[scale = 0.3]{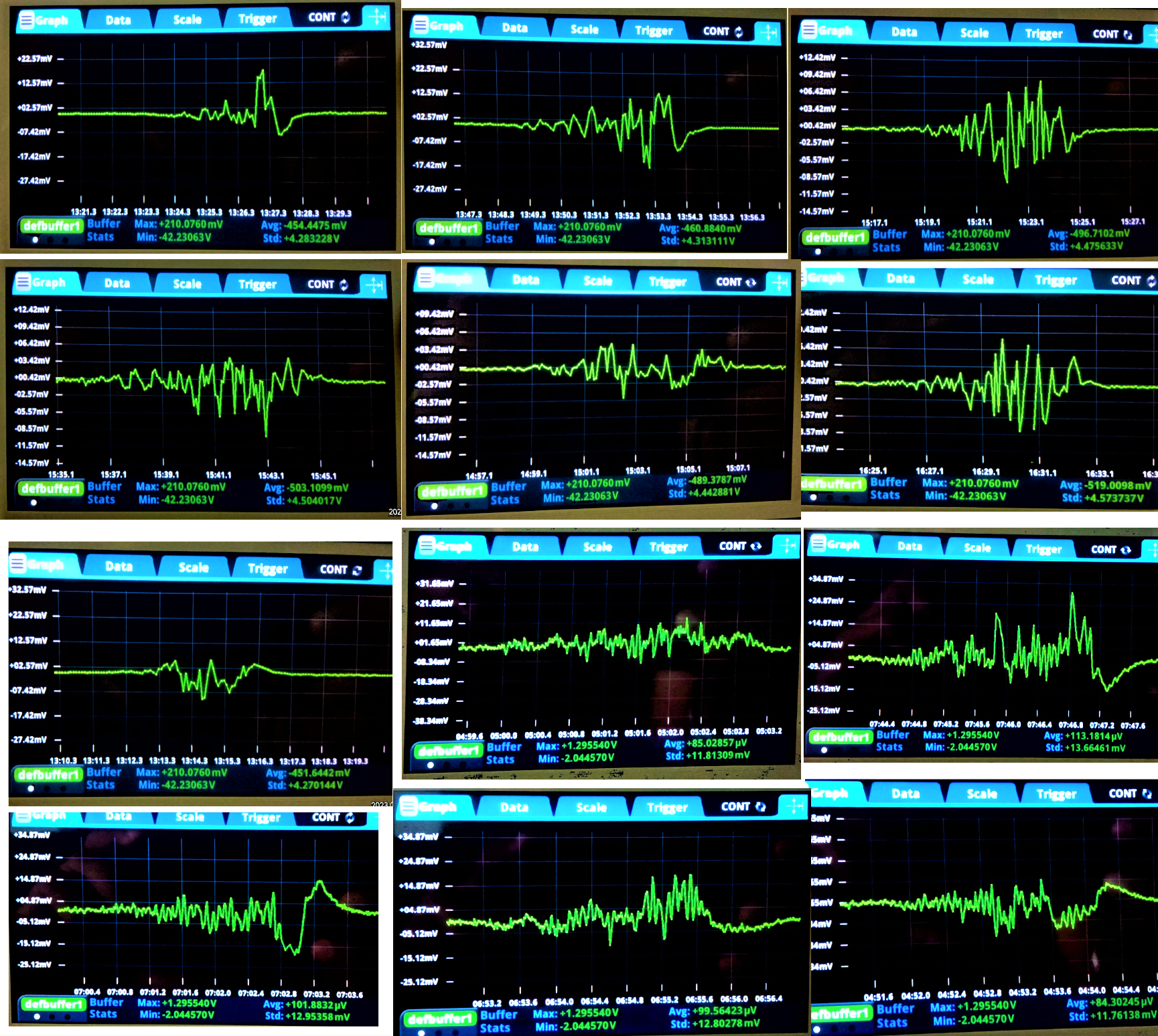}
    \caption{Barkahusen Avalanche waveforms}
    \label{fig:result_image}
\end{figure}

\begin{figure}[h!]
    \centering
    \includegraphics[scale = 0.25]{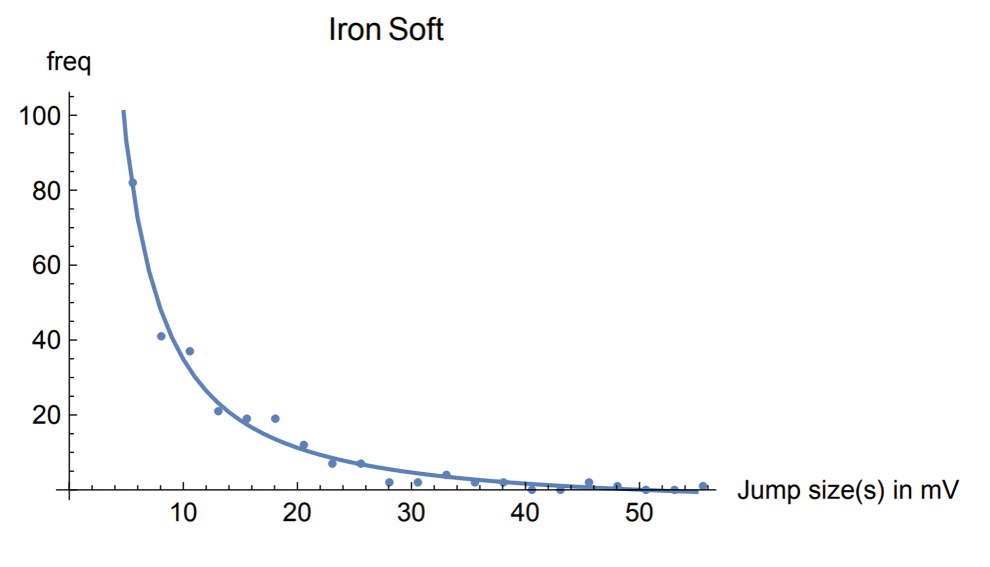}
    \caption{Curve fitting of observed Barkhausen data.}
    \label{fig:plot_fit}
\end{figure}

The recorded waveforms are visually presented in Figure \ref{fig:result_image}.
Upon bringing the magnet close to the sample, the resulting signals are observed in millivolt range, as depicted in the above figure. This indicates the effective performance of our kit, adept at amplifying the microvolt Barkhausen signal into the millivolt range. Subsequently, we conducted measurements to determine the jump size (s) and the frequency (P(s)) of the avalanches. The tail part of the measured data was plotted and fitted with a power-law model (refer to Figure \ref{fig:plot_fit}) given by $P(s) = as^{-\tau} + b$, where 'a' and 'b' represent scaling constants. The obtained $\tau$ value is 1.303, with a=799.139 and b=-4.834.

\section{Statistics of Barkhausen Noise}
In this section, we will address how does our measured value of $\tau$ = 1.303 compare with the available data.\\
In nature, numerous systems exhibit crackling noise when driven slowly, characterized by a series of sudden avalanche-like events with a broad range of sizes and distributions. Various studies have been conducted to comprehend the dynamics behind these phenomena, revealing strikingly similar behavior across diverse systems such as earthquakes, plastic deformation, microfractures, vortices in superconductors, dimming events in stars, and Barkhausen noise \cite{thinFilmm}.

Extensive theoretical and experimental investigations, framed within the Langevian theory of domain wall motion, have contributed to the interpretation of Barkhausen noise statistics. These studies have revealed a power law or scaling law distribution aligning seamlessly with experimental data \cite{Bark_stat_1, Bark_stat_2}.
\begin{equation}
    P(s) \sim s^{-\tau}\\
\end{equation}
\begin{equation}
   P(T) \sim s^{-\alpha} 
\end{equation}

where, \textit{s} is the jump size and \textit{T} is the avalanche duration, 
Durin and Zapperi \cite{Bark_stat_3} conducted an investigation into the scaling properties of Barkhausen Noise across several ferromagnetic materials. They measured avalanche distributions, determined scaling constants, and grouped the materials into two classes with $\tau$ values of $\tau$ = 1.50 $\pm$ 0.05 and $\tau$ = 1.27 $\pm$ 0.03. The first category encompasses Si-Fe polycrystals and the partially crystallized amorphous alloy, while the second category comprises amorphous alloys subjected to stress. These values agree reasonably well with our measured value of $\tau$. \\

%\begin{figure}
    %\centering
   % \includegraphics[scale = 1]{Kappa.png}
   % \caption{Distributions (shifted vertically for clarity) of the Barkhausen jump sizes s = %$|\Delta m_x|$ for different disorder strengths.\cite{Bark_stat_4}}
   % \label{fig:samiKappa}
%\end{figure}

\section{Discussion}
Our designed kit serves as a valuable resource for universities and colleges, facilitating the demonstration of the Barkhausen Effect. Students can easily perceive the Barkhausen noise through the integrated speaker. Not only is the kit simple to construct, but it is also highly cost-effective, requiring a modest investment of approximately Rs. 300 (approximately \$3.61 USD). This accessible tool fosters a deeper understanding among students regarding domain walls and the magnetization dynamics of ferromagnetic materials. Additionally, it proves useful for non-destructive testing of materials to identify impurities.\\

Moreover, the kit enables the calculation of the scaling constant ($\tau$) for the soft iron core, adding an intriguing dimension to its educational utility. The versatility of this kit extends to potential modifications for demonstrating the Barkhausen Effect in various materials. However, it is important to note that the system is susceptible to errors such as mechanical vibrations, magnetic field fluctuations, and environmental noise. While data obtained from this device can serve for preliminary calculations, it is acknowledged that more sophisticated and sensitive instruments would be required for detailed analyses. Nevertheless, at the college or school level, this kit proves to be a valuable resource for effective demonstration purposes.

\section{Conclusion}
In conclusion, the Barkhausen effect is a great educational and research topic. To advance in this field, a foundational understanding is crucial, and this can be achieved through the use of a readily designed kit. Such a kit provides an accessible means for students to delve into the intricacies of the Barkhausen effect, offering valuable insights. The kit is very cost effective and can introduced in colleges and universities in India and abroad. Moreover, with the aid of this straightforward kit, one can effectively calculate the scaling exponent $\tau$ value of a Soft Iron sample and other samples too and can be compared with simulations and other other available data.
\subsection{Acknowledgments}
I would like to thank \textit{Physical Research Laboratory (PRL) -Ahmadabad} and \textit{Indian Institute of Space Science and Technology (IIST) - Kerala} for providing me all the required facilities.

%%%%% References %%%%%

%\bibliography{ref}   % bibliography data in report.bib
%\bibliographystyle{spiejour}   % makes bibtex use spiejour.bst

%%%%% Biographies of authors %%%%%
\end{spacing}
\end{document}